# Semi-Leptonic Decays of $B$-Mesons

$UKQCD$ $Collaboration$ - presented by Laurent Lellouch [a][*]

[a]CPT, CNRS Luminy, Case 907, F-13288 Marseille, France

Form factors relevant for semi-leptonic $B \to D$, $D^*$ decays are obtained in quenched lattice QCD. Tests of heavy-quark symmetry as well as determinations of the Isgur-Wise function and $|V_{cb}|$ are presented.

## 1. Introduction and Simulation Details

The study of semi-leptonic $B \to D$, $D^*$ decays is interesting phenomenologically because it enables the determination of the CKM matrix element $V_{cb}$ and theoretically, because it permits one to test Heavy-Quark Symmetry (HQS) [1]. The matrix elements required to described these decays can be parametrized in terms of six form factors:

$$\frac{\langle D(v')|\bar c\gamma^\mu b|B(v)\rangle}{\sqrt{m_B m_D}} = (v+v')^\mu\, h^+(\omega)$$
$$+(v-v')^\mu\, h^-(\omega)\,, \qquad (1)$$

$$\frac{\langle D^*(v',\epsilon)|\bar c\gamma^\mu b|B(v)\rangle}{\sqrt{m_B m_{D^*}}} = i\epsilon^{\mu\nu\alpha\beta}\epsilon^*_\nu v'_\alpha v_\beta\, h_V(\omega)\,,$$

$$\frac{\langle D^*(v',\epsilon)|\bar c\gamma^\mu\gamma^5 b|B(v)\rangle}{\sqrt{m_B m_{D^*}}} = (\omega+1)\epsilon^{*\mu}\, h_{A_1}(\omega)$$
$$-\epsilon^*\cdot v\,(v^\mu h_{A_2} + v'^\mu h_{A_3})\,,$$

where $\omega = v\cdot v'$ and $\epsilon^\mu$ is the polarization vector of the $D^*$. In the limit that $\Lambda_{QCD}$ is negligible compared to the masses of the $b$ and $c$ quarks, HQS reduces these six form factors to a single universal function of the recoil, $\xi(\omega)$, known as the Isgur-Wise function and normalized to 1 at $\omega = 1$ [1]. Thus, we have

$$h_i(\omega) = (\alpha_i + \beta_i(\omega) + \gamma_i(\omega))\ \xi(\omega)\,, \qquad (2)$$

where $\alpha_+ = \alpha_V = \alpha_{A_1} = \alpha_{A_3} = 1$ and $\alpha_- = \alpha_2 = 0$. The functions $\beta_i$ parametrize perturbative, radiative corrections to the symmetry limit which we calculate using the results of [2]. The functions $\gamma_i$ parametrize non-perturbative corrections which correspond to matrix elements of higher-dimension operators in Heavy-Quark Effective Theory and which are proportional to inverse powers of the heavy-quark masses. One important piece of information is that Luke's theorem [3] guarantees that $\gamma_{+,A_1}(1)$ start at order $(\bar\Lambda/2m_{b,c})^2$, where $\bar\Lambda$ is the energy carried by the light degrees of freedom in the mesons and is typically on the order of 500 MeV [4].

The results presented below were obtained from 60 quenched configurations at $\beta = 6.2$ on a $24^3 \times 48$ lattice. Quark propagators were generated from an $\mathcal{O}(a)$-improved Wilson action [5] at three values of the light-quark hopping parameter around that of the strange (0.14144, 0.14226 and 0.14262) and four values of the initial and final heavy-quark hopping parameter around that of the charm (0.121, 0.125, 0.129, 0.133). Where necessary, results were extrapolated linearly in light-quark mass to the chiral limit ($\kappa_c = 0.14315(2)$). The form factors were obtained from a ratio of three to two-point functions.

## 2. Non-Perturbative Subtraction of Discretization Errors

In conjunction with an appropriate rotation of the quark fields, the use of an $\mathcal{O}(a)$-improved Wilson action guarantees that the leading discretization errors in matrix elements are reduced from $\mathcal{O}(a)$ to $\mathcal{O}(\alpha_s a)$ and $\mathcal{O}(a^2)$ [6]. Nevertheless, because we are working with heavy quarks whose bare masses can be as large as half the inverse lattice spacing, we must worry about discretization errors. We have found them to be as large as 10 to 15% for the most massive of our heavy-quarks. Fortunately, though, they can be at least partially subtracted.

Discretization errors made in evaluating the continuum form factor, $h_+$, can be parametrized

---

[*]This research was supported by the UK Science and Engineering Research Council under grants GR/G 32779 and GR/J 21347. I thank my colleagues from the UKQCD Collaboration and G. Martinelli for fruitful discussions.

in terms of a function $d_+(\omega)$ defined by

$$h_+(\omega) = Z_V(\alpha_s)(1 - d_+(\omega)) h_+^{latt}(\omega) \quad (3)$$

where $h_+^{latt}$ is the form factor obtained from the lattice calculation and $Z_V$ is the usual perturbative renormalization constant. Unlike $Z_V$, the discretization errors, $d_+(\omega)$, are non-perturbative and depend on the initial and final states. To subtract them, we consider the ratio

$$\bar{h}_+(\omega) \equiv (1 + \beta_+(1)) \frac{h_+(\omega)_{latt}}{h_+(1)_{latt}} \quad (4)$$

which, to leading order in $\beta_+$, $\gamma_+$ and $d_+$, is equal to $(1 - \gamma_+(1) + d_+(\omega) - d_+(1)) h_+(\omega)$. Since Luke's theorem suggests that $\gamma_+(1)$ should be no larger than a few percent, $\bar{h}_+$ will be the continuum form factor $h^+$ to good accuracy as long as $d_+(\omega) - d_+(1)$ remains small. As we will see shortly, this appears to be the case.

To subtract discretization errors in our evaluation of $h_{A_1}$ we use an analogous procedure; this form factor is also protected by Luke's theorem at zero recoil.

### 3. Testing Heavy-Quark Symmetry

As mentioned above, our calculations are performed with heavy-quark masses around that of the charm. Therefore, to extrapolate the different form factors to the heavy-quark limit or to a regime relevant for $B$-decays, we must understand how they depend on heavy-quark mass. For this purpose, we study the quantity $\bar{h}_+/(1 + \beta_+)$. To leading order in all corrections and errors, this quantity is equal to $(1 + \gamma_+(\omega) - \gamma_+(1) + d_+(\omega) - d_+(1)) \xi(\omega)$. We plot it in Fig. 1 for four different values of the heavy-quark mass for transitions where initial and final heavy quarks are degenerate in mass. The fact that all four sets of points lie very much on the same curve indicates that the combination of power corrections and discretization errors, $\gamma_+(\omega) - \gamma_+(1) + d_+(\omega) - d_+(1)$, is small over quite a large range of recoils. The smallness of this quantity can mean many different things but the fact that it remains small for a wide range of recoils seems to indicate that both power corrections and discretization errors are small. In any event, it implies that the quantity $\bar{h}_+/(1 + \beta_+)$ depends very little on heavy-quark mass which in turn means that this ratio

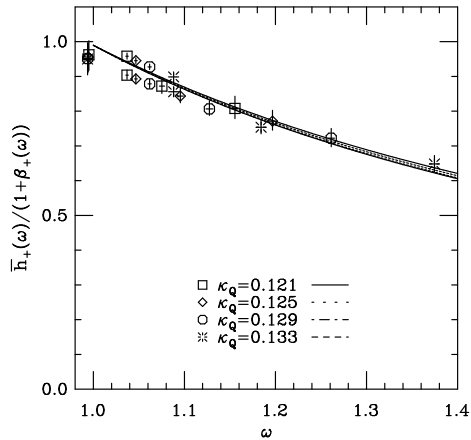

Figure 1. $\bar{h}_+/(1 + \beta_+)$ for four sets of heavy-quark mass. Each set is individually fitted to the parametrization of Eq. (5) (curves). The hopping parameter of the light antiquark is $\kappa_q$=0.14144.

is, to a very good approximation, the Isgur-Wise function.

An identical analysis can be performed for the form factor $h_{A_1}$. We have studied the mass dependence of the quantity $\bar{h}_{A_1}/(1 + \beta_{A_1})$ and have found it to be entirely negligible in the same range of heavy-quark masses and recoils as for $\bar{h}_+/(1 + \beta_+)$. This means that $\bar{h}_{A_1}/(1 + \beta_{A_1})$ should be the same Isgur-Wise function as the one obtained from $\bar{h}_+/(1 + \beta_+)$. This can be seen explicitly in Fig. 2 where we plot the ratio of $(\bar{h}_{A_1}/(1 + \beta_{A_1}))/(\bar{h}_+/(1 + \beta_+))$ for the same degenerate transitions as in Fig. 1.

The near equality of $\bar{h}_+/(1+\beta_+)$ and $\bar{h}_{A_1}/(1+\beta_{A_1})$ and the apparent independence of both these quantities on heavy-quark mass suggests that both flavor and spin components of HQS are well satisfied here even for heavy quarks with masses around that of the charm as long as the Luke-suppressed, $\omega$=1 power corrections which are subtracted by our normalization procedure are not surprisingly large. This is in stark contrast with the symmetry violations found in leptonic decays of $D$ mesons which are on the order of 30 to 40% [7].

### 4. The Isgur-Wise Function and $V_{cb}$

Having established that the quantities $\bar{h}_+/(1+\beta_+)$ and $\bar{h}_{A_1}/(1+\beta_{A_1})$ are, to a good approxima-tion, independent of heavy-quark mass, we can combine the results we have for different initial and final heavy-quark-mass, extrapolate them in light-quark mass to the chiral limit and obtain in this way the Isgur-Wise function, $\xi(\omega)$, relevant for semi-leptonic $B \to D, D^*$ decays. Because our results for $\bar{h}_+/(1+\beta_+)$ are more accurate, we use that quantity alone to determine the function $\xi(\omega)$. We fit this function to $s\xi_{NR}(\omega)$ where $s$ is introduced to absorb uncertainties in overall normalization and $\xi_{NR}(\omega)$ is a standard form for the Isgur-Wise function

$$\xi_{NR}(\omega) = \frac{2}{\omega+1}\exp\left[-(2\rho^2-1)\frac{\omega-1}{\omega+1}\right] \quad (5)$$

which parametrizes $\xi$ in terms of the slope parameter $\rho^2 = -\xi'(1)$. We find

$$\rho^2 = 0.9 \,{}^{+\,2}_{-\,3}\,{}^{+\,4}_{-\,2}, \quad (6)$$

and $s=0.96 \,{}^{+\,2}_{-\,2}\,{}^{+\,5}_{-\,3}$, where the first errors are statistical and the second systematic. The systematic errors, which are meant to measure possible momentum-dependent discretization errors, are determined by fitting the data corresponding to a same initial-final momentum pair but all available heavy-quark mass combinations. This result for $\rho^2$ is in good agreement with earlier propagating-heavy-quark, lattice results [8].

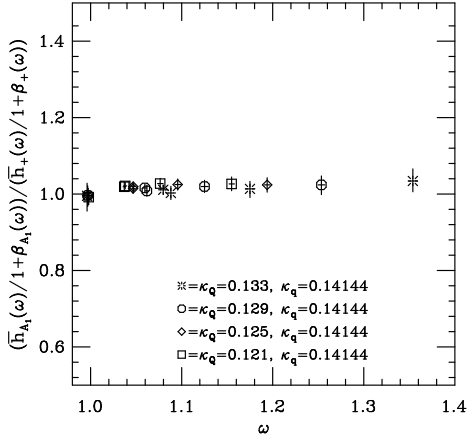

Figure 2. Ratio $(\bar{h}_+/(1+\beta_+)/\bar{h}_{A_1}/(1+\beta_{A_1}))$. The hopping parameter of the light antiquark is $\kappa_q=0.14144$.

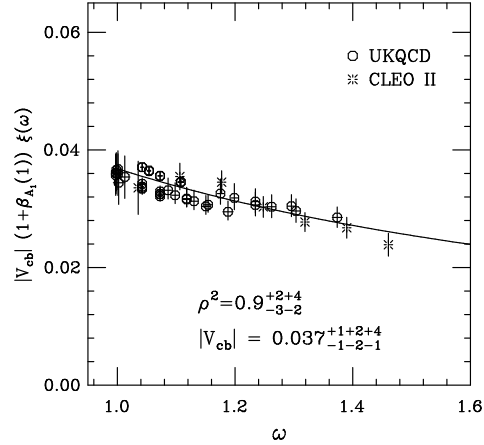

Figure 3. $|V_{cb}|\mathcal{F}(\omega)$ from CLEO (bursts) compared to our result for the Isgur-Wise function rescaled by $|V_{cb}|$ (circles) (see text).

In Fig. 3 we compare our Isgur-Wise function with $|V_{cb}|\mathcal{F}(\omega)$ measured by CLEO [9]. $\mathcal{F}(\omega)$ is the Isgur-Wise function up to radiative and power corrections. If we neglect radiative corrections away from $\omega=1$ and power corrections altogether, a fit of the CLEO data to $|V_{cb}|\xi_{NR}(\omega)$ with $\rho^2$ constrained to it's lattice value of Eq. (6) yields

$$|V_{cb}| = 0.037 \,{}^{+\,1}_{-\,1}\,{}^{+\,2}_{-\,2}\,{}^{+\,4}_{-\,1}\left(\frac{0.99}{1+\beta_{A_1}(1)}\right), \quad (7)$$

where the first set of errors is due to the experimental uncertainties, the second to the statistical errors in our determination of $\rho^2$ and the third to our systematic errors.